\documentclass[prd,amssymb,amsmath,amsfonts,nofootinbib,reprint,showpacs,longbibliography]{revtex4-1}

\usepackage{simpleicons}
\usepackage{graphicx}
\usepackage{lmodern}
\usepackage{paralist}
\usepackage{amsmath,amssymb}
\usepackage{mathrsfs}
\usepackage{amsfonts}
\usepackage[utf8]{inputenc}
\usepackage[dvipsnames]{xcolor}
\usepackage{url}
\usepackage[breaklinks, plainpages=false, colorlinks=true, anchorcolor=violet, linkcolor=violet, citecolor=cyan, urlcolor=violet, bookmarks=false]{hyperref}
\usepackage{multirow}
\usepackage[normalem]{ulem}
\usepackage{lipsum}

\usepackage{marvosym}
\usepackage{enumerate}
\usepackage{color,soul}
\usepackage{acronym}
\usepackage{appendix}

\usepackage{bm}
\usepackage{color}
\usepackage{commath}
\allowdisplaybreaks
\usepackage{multirow}

\def\c3{c_{\rm{N}^3\rm{LO}}}

\def\alphalm0{\alpha_{\ell m 0}}
\def\taulm0{\tau_{\ell m 0}}
\def\omegalm0{\omega_{\ell m 0}}

\newacro{adm}[ADM]{Arnowitt-Deser-Misner}
\newacro{bbh}[BBH]{binary black hole}
\newacro{edgb}[EdGB]{Einstein–dilaton–Gauss–Bonnet}
\newacro{bh}[BH]{black hole}
\newacroplural{bh}[BHs]{black holes}
\newacro{bhns}[BHNS]{black hole-neutron star}
\newacro{bhpt}[BHPT]{black hole perturbation theory}
\newacro{bns}[BNS]{binary neutron star}
\newacro{bf}[BF]{Bayes' factor}
\newacro{cbc}[CBC]{compact binary coalescence}
\newacro{ce}[CE]{Cosmic Explorer}
\newacro{da}[DA]{data analysis}
\newacro{et}[ET]{Einstein Telescope}
\newacro{eob}[EOB]{Effective-One-Body}
\newacro{eom}[EOM]{equations of motion}
\newacro{fd}[FD]{frequency domain}
\newacro{fft}[FFT]{Fast Fourier transform}
\newacro{gw}[GW]{gravitational-wave}
\newacroplural{gw}[GWs]{gravitational-waves}
\newacro{gr}[GR]{general relativity}
\newacro{grb}[GRB]{gamma-ray burst}
\newacro{grhd}[GRHD]{general-relativistic hydrodynamics}
\newacro{gwosc}[GWOSC]{Gravitational Wave Open Science Center}
\newacro{gwtc1}[GWTC-1]{the first gravitational-wave transients catalog}
\newacro{gsf}[GSF]{Gravitational Self Force}
\newacro{hm}[HM]{Higher mode}
\newacroplural{hm}[HMs]{Higher modes}
\newacro{ifo}[IFO]{interferometer}
\newacro{imr}[IMR]{inspiral-merger-ringdown}
\newacro{lr}[LR]{Light Ring}
\newacro{lso}[LSO]{Last Stable Orbit}
\newacro{lvc}[LVC]{LIGO-Virgo Collaboration}
\newacro{lvk}[LVK]{LIGO-Virgo-Kagra Collaboration}
\newacro{lo}[LO]{leading order}
\newacro{ns}[NS]{neutron star}
\newacroplural{ns}[NSs]{neutron stars}
\newacro{nr}[NR]{numerical relativity}
\newacro{nqc}[NQC]{Next-to-quasicircular corrections}
\newacro{nlo}[NLO]{next-to-leading order}
\newacro{nnlo}[NNLO]{next-to-next-to-leading order}
\newacro{n3lo}[N3LO]{next-to-next-to-next-to-leading order}
\newacro{n4lo}[N3LO]{next-to-next-to-next-to-next-to-leading order}
\newacro{ode}[ODE]{Ordinary Differential Equation}
\newacroplural{ode}[ODEs]{Ordinary Differential Equations}
\newacro{pn}[PN]{post-Newtonian}
\newacro{pm}[PM]{post-Minkowskian}
\newacro{pe}[PE]{parameter estimation}
\newacro{psd}[PSD]{power spectral density}
\newacro{pa}[PA]{post-adiabatic}
\newacro{qnm}[QNM]{quasi-normal mode}
\newacro{qc}[QC]{quasi-circular}
\newacro{snr}[SNR]{signal-to-noise ratio}
\newacro{spa}[SPA]{stationary-phase approximation}
\newacro{sxs}[SXS]{Simulating eXtreme Spacetimes}
\newacro{td}[TD]{time domain}
\newacro{ng}[NG]{Next Generation}
\newacro{sm}[SM]{Supplementary Material}
\newacro{aic}[AIC]{Akaike information criterion }
\newacro{de}[DE]{differential evolution}
\newacro{fti}[FTI]{Flexible theory independent}
\newacro{tiger}[TIGER]{Test Infrastructure for General Relativity}
\newacro{pseob}[pSEOB]{pSEOBNR}
\newacro{ppe}[ppE]{Parametrised post-Einsteinian framework}
\newacro{bgr}[bGR]{beyond-GR}
\newacro{kde}[KDE]{kernel density estimator}

\definecolor{cyan}{rgb}{0,0.9,0.9}
\definecolor{orange}{rgb}{0.9,0.5,0}
\definecolor{magenta}{rgb}{1,0,1}
\definecolor{purple}{rgb}{0.8,0.4,0.8}
\definecolor{gray}{rgb}{0.8242,0.8242,0.8242}
\definecolor{dodgerblue}{rgb}{0.12, 0.56, 1.0}

\begin{document}

\preprint{AAPM/123-QED}

\title{Testing General Relativity with GWTC-4.0 through mixture models}

\author{Koustav Chandra}
    \email{koustav.chandra@aei.mpg.de}
    \affiliation{Max Planck Institute for Gravitational Physics (Albert Einstein Institute), Am M\"uhlenberg 1, 14476 Potsdam, Germany}
\author{Juan Calderón Bustillo}%
    \email{juan.calderon.bustillo@gmail.com}
    \affiliation{Instituto Galego de F\'isica de Altas Enerx\'ias, Universidade de Santiago de Compostela, 15782 Santiago de Compostela, Galicia, Spain}
    \affiliation{Department of Physics, The Chinese University of Hong Kong, Shatin, New Territories, Hong Kong}

\date{\today}

\begin{abstract}

Gravitational-wave observations of compact binary mergers have enabled precision tests of gravity in the strong-field dynamical regime. Current approaches combine single-event results that assume deviations from General Relativity (GR) are uniformly distributed across events, limiting their flexibility and potentially biasing the inferred evidence. We introduce a simple mixture-model framework in which a fraction $\zeta$ of events is consistent with GR, while a fraction $1 - \zeta$ deviates from it, without imposing constraints on the population distribution of the deviation parameters. We apply this method to publicly available results from the LIGO-Virgo-KAGRA (LVK) collaboration on O1-O4a compact binary mergers obtained through FTI, pSEOBNR, and KerrPostMerger tests. We find that the data are consistent with all events satisfying GR. However, we obtain respective Bayes factors $\mathcal{B}^{\zeta=1}_{\zeta \neq 1} \simeq 20 $, $10$ and $15$, which are much smaller than those inferred from existing LVK analyses, indicating that the data require greater flexibility in modelling possible deviations than standard approaches permit. In light of our results, we recommend using flexible mixture models to test GR across compact-merger catalogues, unless there are obvious physical motivations to impose more restrictive models, as in the case of graviton-mass estimates.
\end{abstract}

\maketitle
\section{Introduction}

The detection of \acp{gw} from compact binary mergers has made it possible to probe \ac{gr} in the most extreme regime through a large variety of tests that look for deviations from \ac{gr} in individual \ac{gw} events. These tests include parametrised tests~\citep{Yunes:2009ke, Mishra:2010tp, Cornish:2011ys, Li:2011cg, Pai:2012mv, Agathos:2013upa, LIGOScientific:2016lio, Brito:2018rfr, LIGOScientific:2018dkp, LIGOScientific:2019fpa, LIGOScientific:2020tif, Ghosh:2021mrv, LIGOScientific:2021sio, Saleem:2021nsb, Maggio:2022hre, Mehta:2022pcn, Perkins:2022fhr, Puecher:2022sfm, Sanger:2024axs, Chandra:2025jfc, Chiaramello:2025bhi, Pompili:2025cdc, Roy:2025gzv, Grimaldi:2026prn, LIGOScientific:2026fcf}, consistency tests~\citep{Hughes:2004vw, Ghosh:2016qgn, Ghosh:2017gfp, LIGOScientific:2019fpa, Cornish:2014kda, Cornish:2020dwh, Dideron:2022tap, Dideron:2026uis}, polarisation tests~\citep{Wong:2021cmp}, tests of the no-hair theorem~\citep{Carullo:2019flw, Isi:2021iql, Chandra:2025ipu} and searches for hidden asymmetries \citep{CaldernBustillo2025,vgwO4a}. The primary output of most of these tests is commonly a posterior distribution over a series of parameters $\boldsymbol{\delta}$ that measures the level of disagreement between the observed \ac{gw} signal and \ac{gr} predictions, commonly characterised by $\boldsymbol{\delta} = 0$.

The growing catalogue of \ac{gw} events naturally motivates combining information from multiple events to search for deviations from \ac{gr}~\citep{LIGOScientific:2025slb, Nitz:2021zwj, Wadekar:2023gea}. Within the \ac{lvk} collaboration, this is commonly done using two overarching approaches that differ in their assumptions about how deviation parameters are distributed across events. The first constructs a ``joint posterior distribution'' \(p_{\rm joint} (\boldsymbol{\delta}|\{d\})\) by simply multiplying the marginalized posterior probability distribution \(p_k(\boldsymbol{\delta}|d)\) for each event~\citep{LIGOScientific:2021sio}. This, by construction, assumes that the deviation parameter $\boldsymbol{\delta}$ should have a common value across all events. The second, known as the hierarchical approach, relaxes this assumption by treating the deviation parameters as random variables drawn from a population distribution governed by hyperparameters \(\boldsymbol{\Lambda}\)~\citep{Zimmerman:2019wzo}. At present, under the minimum information assumption, this population is modeled as a (multivariate) Gaussian distribution with mean \(\boldsymbol{\mu}\) and covariance \(\boldsymbol{\Sigma}\)~\citep{Isi:2019asy, Zhong:2024pwb}. In this sense, the joint posterior framework can be considered as the $\sigma = 0$ limit of the second.

The above frameworks are physically well-motivated under specific assumptions regarding how deviations from \ac{gr} manifest across a population of events. For example, the joint-posterior method is internally consistent as a null test of \ac{gr}: if \ac{gr} is correct, all deviation parameters are expected to be consistent with the common value \(0\). Likewise, for modified-dispersion tests, if gravitational interactions propagate according to a universal non-\ac{gr} dispersion relation, all events should be consistent with the same underlying dispersion parameter governing the propagation law. In such cases, combining events under the assumption of a common deviation parameter is physically well motivated.

However, in general, the fractional deviation parameters employed are theory-agnostic. Consequently, there is no reason to expect that all events share the same deviation value. In fact, most \ac{bgr} theories predict that deviations depend on intrinsic properties of the sources, often in highly non-linear and theory-specific ways~\citep{Berti:2015itd, Julie:2019sab, Yunes:2024lzm}. Therefore, enforcing a common deviation parameter across the entire catalogue is overly restrictive.

The hierarchical framework makes a less restrictive assumption by promoting the deviation parameters to random variables drawn from a population distribution characterised by certain hyperparameters. In its current form, such a distribution is assumed to be Gaussian and independent of the source's intrinsic properties. 

In addition, both of the above frameworks implicitly impose the additional restriction that sources cannot coexist when some satisfy GR and others deviate from it, or from any alternative baseline model. Such an assumption is not appropriate if one considers, e.g., the existence of exotic subfamilies of compact objects such as boson stars \cite{Kaup1968, Brito:2018rfr, CalderonBustillo:2020fyi}. 
In this situation, enforcing a single shared population model can dilute the signatures of such alternatives in the combined inference. As a consequence, catalogue-level analyses based on these frameworks may systematically underestimate the evidence for heterogeneous or source-dependent departures from \ac{gr}.\\

In this work, we introduce a simple Bayesian framework based on mixture models that provides maximal flexibility to accommodate deviations from \ac{gr} across a set of events. The proposed approach allows for the possibility that a finite fraction of events violate \ac{gr} and quantifies the corresponding statistical evidence. The framework uses a simple mixture model characterised by a single parameter \(\zeta \in [0,1]\), which represents the fraction of events consistent with \ac{gr}; \(\zeta=1\) indicates that all events satisfy \ac{gr}. While the framework can incorporate specific assumptions regarding the population distribution of deviations, including those adopted by existing methodologies, it remains fully agnostic in its default form. As we show in Appendix~\ref{app:compare}, the two current frameworks used by the \ac{lvk} are restricted versions of this mixture model. In all three cases, the hypothesis that the event catalogue is consistent with \ac{gr} is defined by the same models, with different parameter restrictions imposed on the catalogue model in each framework.

For each test, the only required input is the set of Bayes factors against \ac{gr} obtained from individual event analyses. With this, we compute the posterior distribution \(p(\zeta \mid \{d\})\) together with the Bayesian evidence for a pure \ac{gr} model constrained to \(\zeta=1\) and a more general model with freely running \(\zeta\). With this, we compute the corresponding catalogue-level Bayes Factor in favour of GR as ${\cal{B}}^{\rm GR}_{\rm bGR} = {\cal{B}}^{\zeta = 1}_{\zeta \neq 1}$. 

Applying this framework to selected tests of \ac{gr} performed by the \ac{lvk} on O1-O4a events, we obtain Bayes factors of $\mathcal{O}(10-20)$ in favour of \ac{gr}. Crucially, these Bayes factors are much lower than those derived from existing catalogue-level analyses of \ac{gw} signals performed by the \ac{lvk}, which underscores that the extra freedom our mixture models offer is necessary to accommodate existing events within a beyond-\ac{gr} framework.

The rest of this paper is organised as follows. In Section~\ref{sec:method} we describe our mixture model framework. In Section~\ref{sec:validation}, we demonstrate the method using mock catalogues and demonstrate how multiple individually inconclusive events can collectively provide statistical evidence for deviations from \ac{gr}. Next, in Section~\ref {sec:results}, we apply the framework to three different tests of \ac{gr} conducted by the \ac{lvk} collaboration. In Section~\ref{sec:when}, we show that standard frameworks would be statistically preferred against our flexible model if \ac{gr} deviations satisfied their restrictions. Finally, we conclude this paper with some remarks.

\section{Testing GR through a counting experiment}
\label{sec:method}

Given the strain data $d$ from a \ac{gw} event, the relative support for two competing hypotheses A and B is quantified by the Bayes factor ${}_k\mathcal{B}^{\rm A}_{\rm B} = \mathcal{Z}_{\rm A}/\mathcal{Z}_{\rm B}$, where
\begin{equation}\label{eq:evidence}
    \mathcal{Z} = \int \pi(\boldsymbol{\theta})\,\mathcal{L}(d \mid \boldsymbol{\theta})\,\mathrm{d}\boldsymbol{\theta}
\end{equation}
is the Bayesian evidence, $\mathcal{L}(d \mid \boldsymbol{\theta})$ is the likelihood, and $\pi(\boldsymbol{\theta})$ is the prior. In what follows, model A denotes \ac{gr}, with source parameters $\boldsymbol{\theta}_{\rm GR}$, while model B denotes an alternative \ac{bgr} model with parameters $\boldsymbol{\theta}_{\rm bGR}=\{\boldsymbol{\theta}_{\rm GR}, \boldsymbol{\delta}\}$. Although not required by our formalism, in all cases considered in this work, the \ac{gr} model is nested within the alternative model, with \ac{gr} corresponding to $\boldsymbol{\delta} = 0$ limit of the alternative model. Values $\mathcal{B}^{\rm GR}_{\rm bGR} > 1$ indicate preference for \ac{gr}, while $\mathcal{B}^{\rm GR}_{\rm bGR} < 1$ favour deviations.\footnote{The Bayes factor can be computed either via two independent \acp{pe} runs (under \ac{gr} and \ac{bgr} models) or, more efficiently, through the Savage-Dickey ratio, exploiting the fact that \ac{gr} is nested within the parameterized model at \(\boldsymbol{\delta}=0\) (See Appendix~\ref{sec:savage} for details)}\\

Given two competing models and a catalogue of \(N\) events, we ask whether finite fractions $\zeta \in [0,1]$ and $1-\zeta$ of the events may be better explained by each of the models. Under this framework, the likelihood for a single event is \citep{Callister:2022qwb}:
\begin{align}
    &\mathfrak{L}(d_k \mid \zeta) = \zeta~\mathcal{Z}_{\rm GR}(d_k) + (1-\zeta)~\mathcal{Z}_{\rm bGR}(d_k) \\
    &\implies \frac{\mathfrak{L}(d_k \mid \zeta)}{\mathcal{Z}_{\rm bGR}(d_k)} = \zeta \cdot {}_k\mathcal{B}^{\rm GR}_{\rm bGR} + (1-\zeta) ~.
\end{align}
Assuming events are independent, the likelihood for the whole catalogue is:
\begin{equation}\label{eq:analysed-likelihood}
    \ln \mathfrak{L}(\{d_k\} \mid \zeta) = \sum_k \ln \left(\zeta \cdot {}_k\mathcal{B}^{\rm GR}_{\rm bGR} + (1-\zeta)\right) + \sum_k \ln \mathcal{Z}_{\rm GR}^k~.
\end{equation}
The second term is independent of $\zeta$ and can be dropped when inferring the catalogue-level log posterior:
\begin{equation}
    \ln \mathfrak{p}(\zeta \mid \{d_k\}) \propto  \sum_k \ln \left(\zeta \cdot {}_k\mathcal{B}^{\rm GR}_{\rm bGR} + (1-\zeta)\right) + \ln \pi(\zeta)~.
\end{equation}
We choose \(\pi(\zeta)\) to be uniform between \([0,1]\), with \(\zeta=1\) indicating that the entire analysed population is consistent with \ac{gr}. The \textit{analysed} catalog-level evidence for \ac{gr} is then simply the posterior density at 1: \(\mathfrak{p}(\zeta=1 \mid \{d_k\})\).\\

Using mixture models replaces the ``all-or-nothing'' approach of current overarching methods with a scenario in which two different families of events can coexist. Here, the hypothesis that GR is the correct theory for all events corresponds to the limit $\zeta=1$, and the Bayes factor for GR against alternative models is given by the ratio of the evidences $\mathcal{Z}_{\zeta=1}/\mathcal{Z}_{\zeta \neq 1}$. Conclusive evidence against or in favour of GR can be naturally spotted through the accumulation of small inconclusive evidence across the catalogue, which may lead to $\mathcal{Z}_{\zeta=1}/\mathcal{Z}_{\zeta \neq 1} \ll (\gg) 1$.


\section{Validation and Testing: finding a violation of GR in a synthetic catalog}
\label{sec:validation}

\begin{figure}
    \centering
    \includegraphics[width=0.5\textwidth]{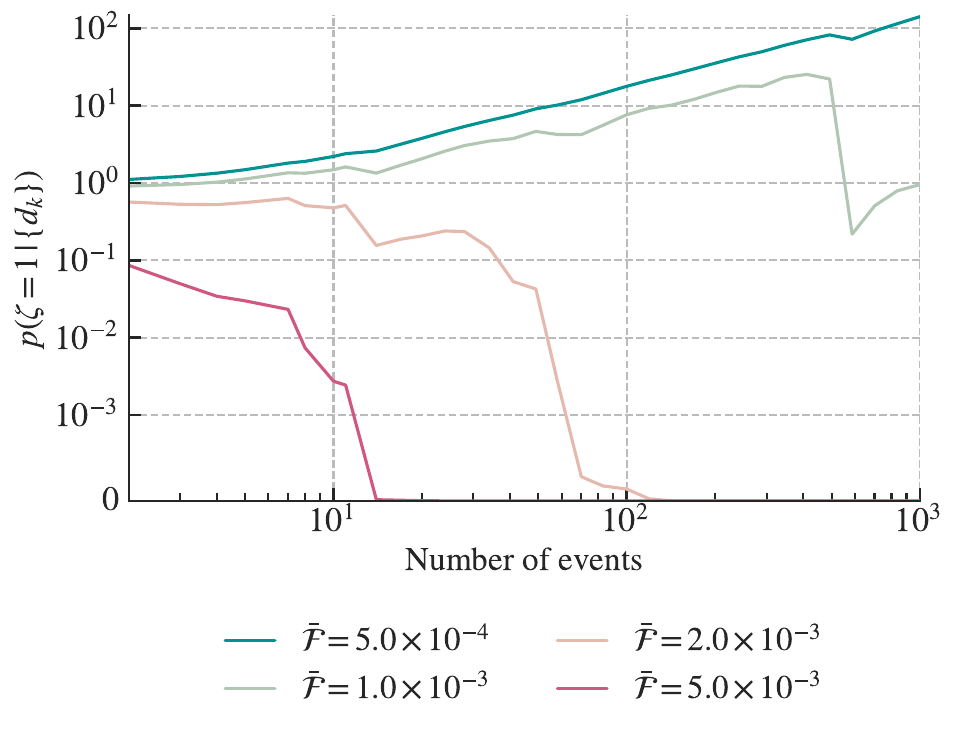}\\
    \caption{Toy model demonstration of the mixture model for different choices of disagreement between the \ac{gr} waveform and \ac{bgr} waveform.}
    \label{fig:toy}
\end{figure}

Before diving into the analysis of GWTC-4 events, we illustrate our method with a toy scenario in which \ac{gr} is not the correct underlying theory. Concretely, we assume toy true signals $h(\theta)+\Delta h$ that differ from the \ac{gr} waveforms that best fits them by some amount $\Delta h$. Specifically, these differences translate into a mismatch \(\bar{\mathcal{F}}\) between our toy waveforms and the \ac{gr} best-fit \ac{gr} templates~\citep{Apostolatos:1995pj}. With this, we can approximate the single-event Bayes factor as:
\begin{equation}\label{eq:crackpot-bayes-factor}
    \ln \mathcal{B}^{\rm GR}_{\rm bGR}
    \approx -\frac{\rho^2 \bar{\mathcal{F}}}{2}
    + \Delta_{\rm Ockham}\,.
\end{equation}
where \(\rho\) is the maximum network \ac{snr} that GR templates can recover and \(\Delta_{\rm Ockham} \equiv \ln \pi_{\rm bGR}(\boldsymbol{\delta}=0)\) is the Ockham penalty favoring the simpler \ac{gr} model. Considering an uniform prior \(\delta \in [-1,1]\), this gives \(\Delta_{\rm Ockham} = \ln(0.5) \approx -0.69\) for a single deviation parameter. The competition between the first and second terms of Eq. \ref{eq:crackpot-bayes-factor} defines a critical \ac{snr},
\begin{equation}
    \rho_{\rm crit} = \sqrt{\frac{2|\Delta_{\rm Ockham}|}{\bar{\mathcal{F}}}}\,,
\end{equation}
below which individual events favor \ac{gr} despite this not being the correct signal model. Below this SNR, \ac{gr} deviations are too weak to compensate for the larger parameter space spanned by non-GR templates.\\

We study the detectability of the beyond GR effects as function of the mismatch \(\bar{\mathcal{F}}\).  We simulate a catalogue of \(1000\) binary merger signals with network \ac{snr} drawn from a power-law distribution \(p(\rho)\propto \rho^{-4}\) over \(\rho \in (20,200) \), consistent with the expected detectability distribution for a uniform-in-Euclidean-volume population. We adopt representative mismatches and compute the per-event Bayes factor using Eq.~\eqref{eq:crackpot-bayes-factor}. \\

Figure~\ref{fig:toy} summarises our results for different values of \(\bar{\mathcal{F}}\), as a function of the number of detected events. Deviations from GR leading to small values of \(\bar{\mathcal{F}}\) require substantially larger catalogues for the posterior support for a purely \ac{gr}-consistent population \( p(\zeta=1 \mid \{d_k\})\) be driven to zero. In contrast, for larger \(\bar{\mathcal{F}}\), a smaller amount of events is sufficient to conclude that the analysed catalogue contains, at least, a subset of \ac{gr}-violating events. 

We note that we observe a drop for \(\bar{\mathcal{F}}=10^{-3}\) at \(\sim 600\) events. This is caused by the random occurrence of a particularly loud event that temporarily dominates the catalogue. As additional \ac{gr}-favoring events are added, however, the evidence if favour of GR keeps growing. Incidentally, this preempts one of the advantages of this framework, namely its robustness against potential signal affected by systematic errors. While such events may lead to apparent conclusive false evidence against GR if considered in isolation~\citep{Pang:2018hjb}, such evidence is naturally washed away when combined with the rest of the catalogue. We discuss this with a more concrete example later.\\


The above results naturally motivate a key question: how many \ac{gr}-violating events are needed to detect true \ac{gr} violations if most events are consistent with \ac{gr}? In Appendix~\ref{sec:violation}, we show that for events with modest individual evidence against \ac{gr}---quantified by an average per-event Bayes factor \(\langle \mathcal{B}^{\rm GR}_{\rm bGR} \rangle \sim 10^{-\beta}\), with \(\beta \geq 0\)---the number of violating events required for \(3\sigma\)-equivalent population-level detection scales as \(N_{\rm violating} \gtrsim 1.4/\beta\). This means that even when individual events are inconclusive (\(\beta \sim 0.3\), corresponding to a Bayes factor of \(\sim 2\) against \ac{gr} per event), detecting a 10\% violating subpopulation in a catalogue of 50 events requires only \(\mathcal{O}(5)\) such events. This demonstrates the power of population-level evidence accumulation: deviations observed in a small fraction of events can produce conclusive results despite being individually undetectable.

\section{Application to Real Gravitational-Wave Events}
\label{sec:results}

We now apply our framework to publicly available results from three different types of parametrised tests of \ac{gr} performed by the \ac{lvk} Collaboration using events from the O1--O4a observing runs: the \ac{fti}~\citep{Mehta:2022pcn}, pSEOBNR~\citep{Brito:2018rfr, Ghosh:2021mrv, Maggio:2022hre, Pompili:2025cdc, Grimaldi:2026prn}, and pyRing--KerrPostMerger (TEOBPM)~\citep{Gennari:2023gmx} analyses. The corresponding catalogue-level results are reported in~\citet{LIGOScientific:2026fcf, LIGOScientific:2026wpt}, while the associated posterior samples and event-level Bayes factors are publicly available in Refs.~\citep{LIGO_P2600130_2026, LIGO_P2600129_2026}.

The \ac{fti} and pSEOBNR analyses probe deviations from \ac{gr} by introducing phenomenological deformations into the inspiral, merger, and ringdown sectors of waveform models constructed within the \textsc{SEOBNRv5} family~\citep{Pompili:2023tna, Ramos-Buades:2023ehm}. In contrast, the TEOBPM analysis focuses specifically on possible post-merger deviations from the Kerr black-hole paradigm, using non-precessing, non-eccentric \textsc{TEOBResumS} waveforms as the \ac{gr} baseline~\citep{Damour:2014yha, DelPozzo:2016kmd}. 

\subsection*{FTI test}
\begin{figure}
    \centering
    \includegraphics[width=0.5\textwidth]{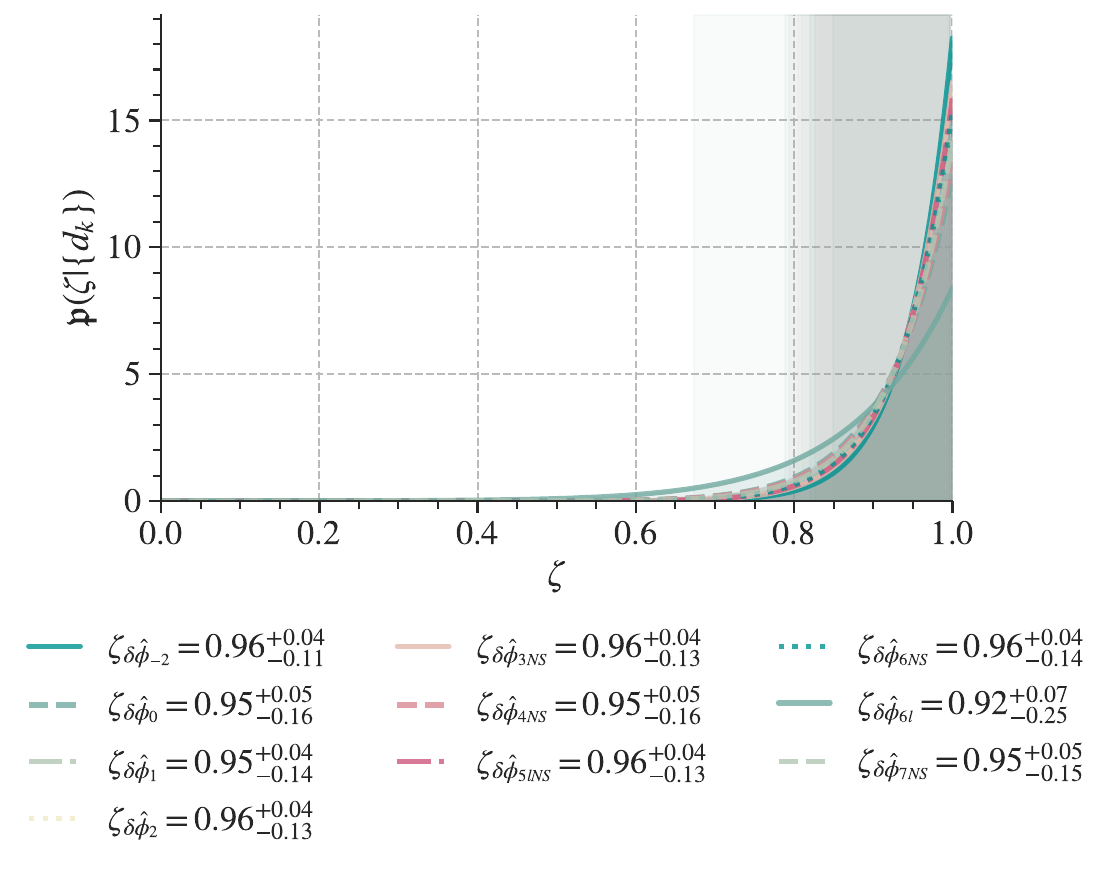}
    \caption{Posterior distributions for the \ac{gr}-consistent fraction \(\zeta\) of \ac{gw} events, inferred using mixture models applied to the \ac{fti} analysis. The different colours and line styles correspond to different deviation parameters, each showing that almost all analysed events are consistent with \ac{gr}.}
    \label{fig:fti}
\end{figure}

The \ac{fti} test probes deviations in the inspiral phase evolution within the non-spinning sector, under the assumption of non-eccentric, non-precessing binaries. To this, the baseline adiabatic inspiral phase is modified by promoting the non-spinning \ac{pn} coefficients as
\begin{equation}
    \psi_i \;\rightarrow\; (1 + \delta \hat{\phi}_i)\,\psi_i^{\rm GR,NS} + \psi_i^{\rm GR,S}\,,
\end{equation}
where \(\delta \hat{\phi}_i\) are fractional deviation parameters, and \(\psi_i^{\rm GR,NS}\) and \(\psi_i^{\rm GR,S}\) denote the non-spinning and spin-dependent \ac{gr} contributions, respectively. For the case of O4a events, the ten deviation parameters
\(\{\delta \hat{\phi}_{-2}, \delta \hat{\phi}_{0}, \delta \hat{\phi}_{1}, \delta \hat{\phi}_{2}, \delta \hat{\phi}_{3}, \delta \hat{\phi}_{4}, \delta \hat{\phi}_{5l}, \delta \hat{\phi}_{6}, \delta \hat{\phi}_{6l}, \delta \hat{\phi}_{7}\}\)
are sampled independently, without assuming any specific beyond-\ac{gr} theory (here, the sub-index \(l\) denotes logarithmic contributions). For these cases, and for each parameter $\delta \hat{\phi}_{j}$ we compute the individual-event evidence $\mathcal{B}^{\rm GR}_{\rm bGR, (i,j)}$ through the Savage-Dickey density ratio evaluated at $\hat{\phi}_{j}=0$.

Figure~\ref{fig:fti} presents a summary of the catalogue-level results. The \(\zeta\) posteriors consistently rail at unity for all deviation parameters, indicating that the observed event catalogue is largely consistent with all events satisfying \ac{gr}. The catalogue-level Bayes factors, $\mathcal{B}^{ \zeta = 1 }_{\zeta \neq 1, j} \in (5, 20)$, provide moderate/strong support for the scenario in which all events satisfy the baseline GR model. This contrasts sharply with the much larger Bayes Factors (mostly in the very strong/decisive support range) obtained using both \ac{lvk} frameworks, shown in the two rightmost columns of Table \ref{tab:catalog_results}. Although the \ac{lvk} analyses report only posterior distributions computed using the joint and hierarchical frameworks, we estimate the corresponding Bayes Factors by comparing the prior and posterior probability densities(see Appendix~\ref{app:hierarchical_bayes_factors} for details). This gives much higher evidence favouring \ac{gr} than that returned by the present framework. Specifically, the corresponding Bayes factors range from $\simeq (3,10^{4})$ in the hierarchical framework, while the joint-posterior framework yields values in the range $\simeq (36,10^{3})$, both of which generally provide highly conclusive evidence in favour of the baseline \ac{gr} waveform model. 

\begin{table}[t]
\centering
\caption{Catalogue-level constraints on deviations from \ac{gr} for the \ac{fti} test. We report the Bayes factor from the mixture model analysis for a \ac{gr}-consistent population, along with the one obtained via hierarchical and joint. We assume a prior on the deviation parameters $\pi(\delta \hat{\phi}_k)=\mathcal{U}(-20, 20)$ for the joint analysis. We find that the mixture model's Bayes factor is always smaller than the other two.}
\label{tab:catalog_results}
\begin{ruledtabular}
\begin{tabular}{l c c c}
FTI parameter &
\(\log_{10}\frac{p(\zeta=1 \mid \{d_k\})}{\pi(\zeta=1)}\) &  $\log_{10}^{\rm Hier} \mathcal{B}$ & \( \log_{10}^{\rm Joint} \mathcal{B} \) \\
\hline
\(\delta \hat{\varphi}_{-2}\) & 1.261 & 4.30 & 2.96 \\
\(\delta \hat{\varphi}_{0}\)  & 1.107 & 3.05 & 2.83 \\
\(\delta \hat{\varphi}_{1}\)  & 1.162 & 2.17 & 2.38 \\
\(\delta \hat{\varphi}_{2}\)  & 1.194 & 2.65 & 2.59 \\
\(\delta \hat{\varphi}_{3}\)  & 1.217 & 3.15 & 2.93 \\
\(\delta \hat{\varphi}_{4}\)  & 1.127 & 1.71 & 2.19 \\
\(\delta \hat{\varphi}_{5l}\) & 1.200 & 2.80 & 2.73 \\
\(\delta \hat{\varphi}_{6}\)  & 1.187 & 2.40 & 2.63 \\
\(\delta \hat{\varphi}_{6l}\) & 0.923 & 0.48 & 1.54 \\
\(\delta \hat{\varphi}_{7}\)  & 1.150 & 1.93 & 2.31 \\
\end{tabular}
\end{ruledtabular}
\end{table}

\begin{figure}
    \centering
    \includegraphics[width=0.5\textwidth]{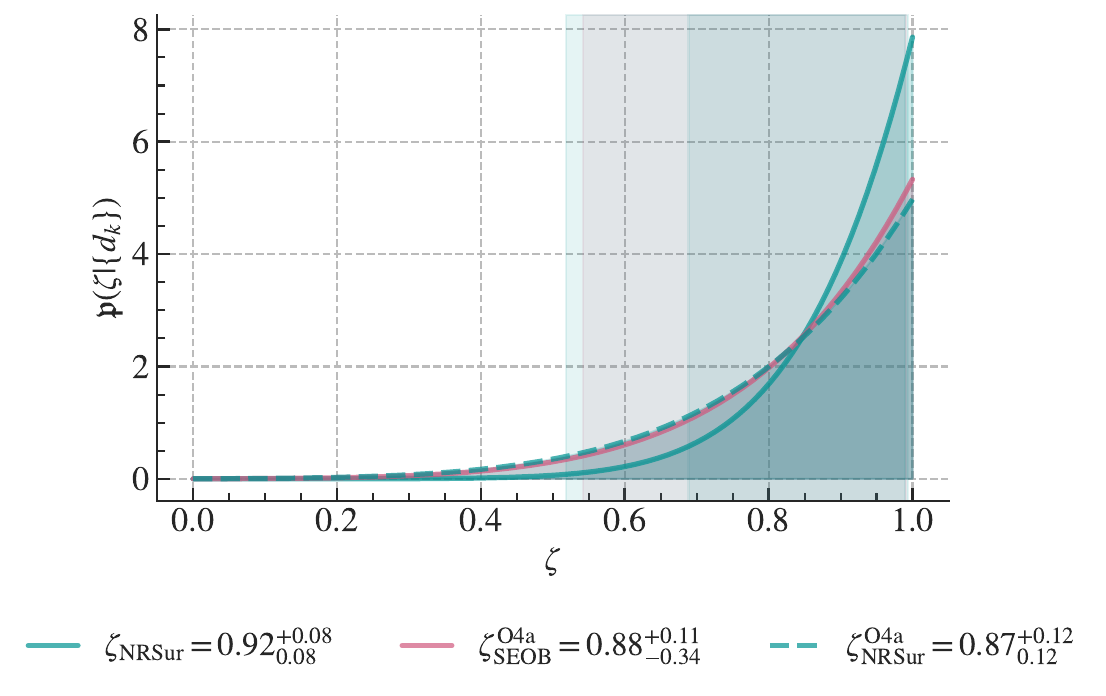}
    \caption{Posterior distributions for the fraction, \(\zeta\), of \ac{gw} events consistent with \ac{gr}, inferred using the mixture-model framework applied to pSEOBNR analyses. The solid green curve uses all events from O1--O4a that were analysed with both pSEOBNR and NRSur, while the dashed green curve is restricted to the corresponding common-event subset from O4a. The dashed red curve shows the result obtained using only the O4a events when comparing against SEOBNR. In all cases, the posterior accumulates near \(\zeta=1\), providing no evidence for a subpopulation of events inconsistent with \ac{gr}.}
    \label{fig:pseob}
\end{figure}
\subsection*{pSEOBNR test}

The pSEOBNR analysis tests the consistency of the post-inspiral signal with those expected from a \ac{gr}-consistent inspiral. It achieves this by introducing controlled fractional deviations $\delta\hat{f}_{\ell m 0}$ and $\delta\hat{\tau}_{\ell m 0}$ in the frequency and damping time of the predicted dominant \acp{qnm} in the faithful SEOBNR waveform model~\citep{Damour:1997sd, Buonanno:1998gg, Buonanno:2000ef, Damour:2014yha}:
\begin{equation}
  f_{\ell m 0} = f^{\mathrm{GR}}_{\ell m 0}\left(1 + \delta\hat{f}_{\ell m 0}\right),
  \quad
  \tau_{\ell m 0} = \tau^{\mathrm{GR}}_{\ell m 0}\left(1 + \delta\hat{\tau}_{\ell m 0}\right)~.
  \label{eq:pseob_deviations}
\end{equation}
Here the \ac{gr} values, \((f^{\mathrm{GR}}_{\ell m 0}, \tau^{\rm GR}_{\ell m 0})\) depend on the remnant mass and spin inferred from the progenitor binary via \ac{nr} fits~\citep{Jimenez-Forteza:2016oae, Hofmann:2016yih}. The standard pSEOBNR test amounts to comparing this parameterised model against its undeformed SEOBNR baseline; \ac{gr} is recovered when the deviations are consistent with zero. At current \ac{snr}, the analysis only targets deviations in the dominant $(2,2,0)$ mode, sampling jointly over $\bigl(\delta\hat{f}_{220},\,\delta\hat{\tau}_{220}\bigr)$ alongside the standard \ac{bbh} parameters. 

Because NRSur7dq4 waveforms \citep{Varma:2019csw} provide (a priori) a more faithful representation of GR than SEOBNR, we additionally compare the Bayesian evidences returned by pSEOBNR and NRSur7dq4 directly, offering a complementary consistency check that does not rely on the SEOBNR baseline. This choice reduces the risk of beyond-GR parameters capturing actual GR physics not reproduced by the baseline GR model, which could lead to false deviations from GR \cite{Pang:2018hjb}. 




We note, however, that the NRSur7dq4 has a limited waveform length, which prevents analysis of systems with redshifted masses \(\gtrsim 60\,M_\odot\) when the signal analysis is started at \(20\,\rm Hz\). This forces us to exclude the GW170104 event. In addition, publicly available \ac{gr} analyses using the same SEOBNRv5PHM baseline as the parameterised runs are not consistently available, making this hybrid approach necessary. In addition to events from the GWTC-4 catalog, we include the event GW230814~\citep{LIGOScientific:2025cmm}---a loud single-detector event not part of \citet{LIGOScientific:2026wpt} -- and GW191109 -- which previously showed an apparent \ac{gr} violation attributable to noise systematics ---as a further test of the method robustness.

Figure~\ref{fig:pseob} shows the posteriors on \(\zeta\) obtained by each of the three mentioned analyses. Solid red and  dashed green curves correspond to O4a events, respectively using using the SEOBNR and NRSur7dq4 waveform models as our reference GR models. In both cases, the posteriors are consistent with $\zeta = 1$, indicating that the catalogue is fully consistent with \ac{gr}. Both analyses lead to almost identical Bayes factors of $\mathcal{B}^{\zeta =1}_{\zeta \neq 1}\simeq 5$, denoting positive support for GR. 

The green solid line corresponds to the inclusion of several O1-O3 events for which NRSur7dq4 parameter inference runs have been publicly released, either by the LVK or by the analysis of GWTC-3 events published by \citet{Islam2025}. We make use of LVK runs when available and from \citet{Islam2025} otherwise. The inclusion of O1-O3 events increases the evidence for GR across the catalog to $\mathcal{B}^{\zeta =1}_{\zeta \neq 1}\simeq 8$.

Regarding the catalogue-level evidence for GR, we obtain Bayes Factors of 3 and 8 for our two \ac{gr} model choices. Once again, these are much smaller than those derived from current frameworks. In particular, under the hierarchical framework we obtain a Bayes Factor favouring GR of \(>10^{3}\) while under the joint posterior method we obtain a value of $\simeq 25$.

\subsection*{TEOBPM test}

\begin{figure}
    \centering
    \includegraphics[width=0.5\textwidth]{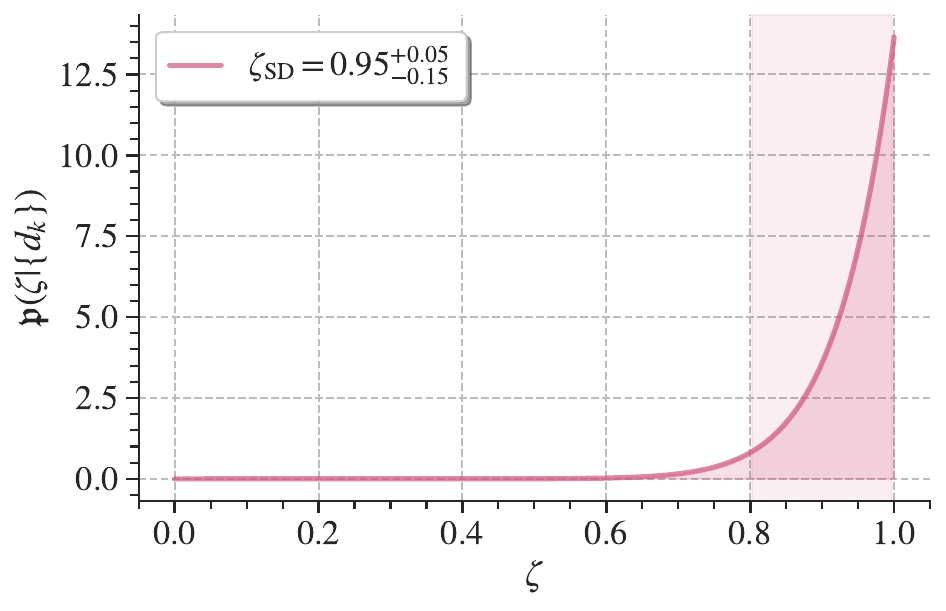}
    \caption{Posterior distribution for the \ac{gr}-consistent fraction \(\zeta\) of \ac{gw} events, inferred using mixture models applied to TEOBPM analysis.}
    \label{fig:pteob}
\end{figure}

The pyRing–KerrPostMerger (TEOBPM) analysis~\citep{Gennari:2023gmx} complements the pSEOBNR test by identifying deviations in the post-merger ringdown spectrum of the remnant black hole, assuming a non-precessing, non-eccentric binary source. Unlike the pSEOBNR test, this framework examines only the data after the peak of the \ac{gw} signal and excludes earlier data, scanning multiple possible starting times. We consider only the analysis results that allow for simultaneous variations in the fundamental \ac{qnm} frequency and damping time, and compute Bayes factors using the Savage–Dickey density ratio at the point in parameter space where all deviations are zero.

Fig.~\ref{fig:pteob} shows that the inferred posterior on \(\zeta\) rails against unity. As in the previous tests, this is consistent with the studied catalog being overwhelmingly \ac{gr}-like. In particular, we obtain a Bayes factor in favour of $\zeta = 1$ -- as opposed to a generic value of $\zeta$ -- of $\mathcal{B}^{\zeta=1}_{\zeta \neq 1} = 13.7$. As in the previous test, this value shows moderate support relative to the \ac{lvk} framework for the hypothesis that the whole catalogue is consistent with GR. For comparison, we find that the hierarchical framework yields a much less conservative value of $10^{2.6}$. We refer the reader to Appendix~\ref{app:hierarchical_bayes_factors} for details. 

\section{When are current frameworks preferred?}
\label{sec:when}
\begin{figure}
    \centering
    \includegraphics[width=0.5\textwidth]{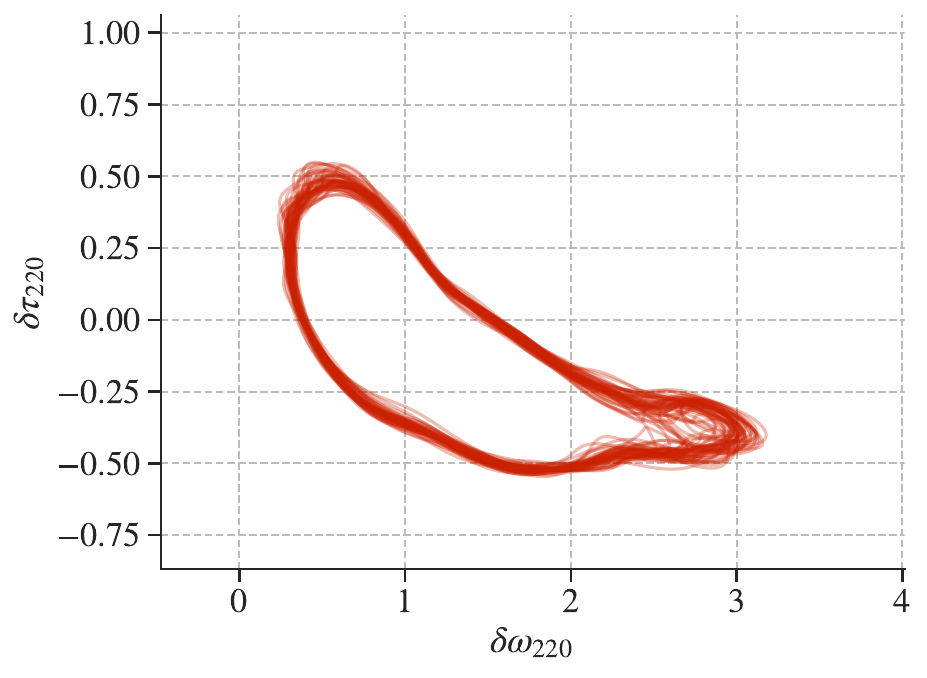}
    \caption{The \(N=50\) copies of GW191109 used in our mock catalogue. Each curve encloses the 90\% credible contour in the \((\delta\omega_{220}, \delta \tau_{220})\) plane obtained from a distinct random draw of 1000 posterior samples from the full pSEOBNR posterior of GW191109. All realisations are centred away from the \ac{gr} value =(0,0), reflecting the apparent violation reported for this event}
    \label{fig:synthetic_events}
\end{figure}

\begin{figure}
    \centering
    \includegraphics[width=0.5\textwidth]{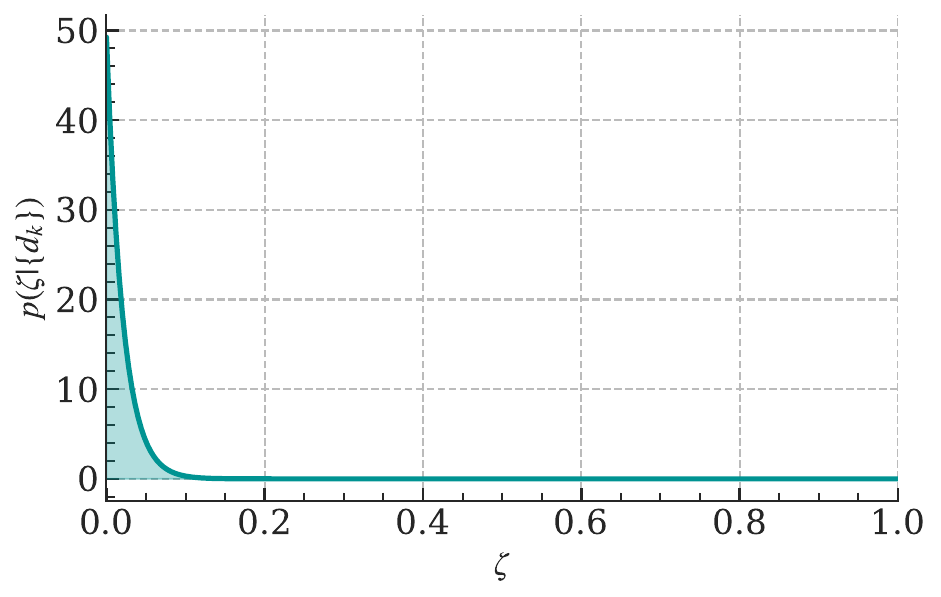}
    \caption{Posterior distribution for the \ac{gr}-consistent fraction \(\zeta\) of \ac{gw} events, inferred using mixture models applied to the mock catalogue of GR-violating signals.}
    \label{fig:mock-mixture}
\end{figure}

\begin{figure}
    \centering
    \includegraphics[width=0.5\textwidth]{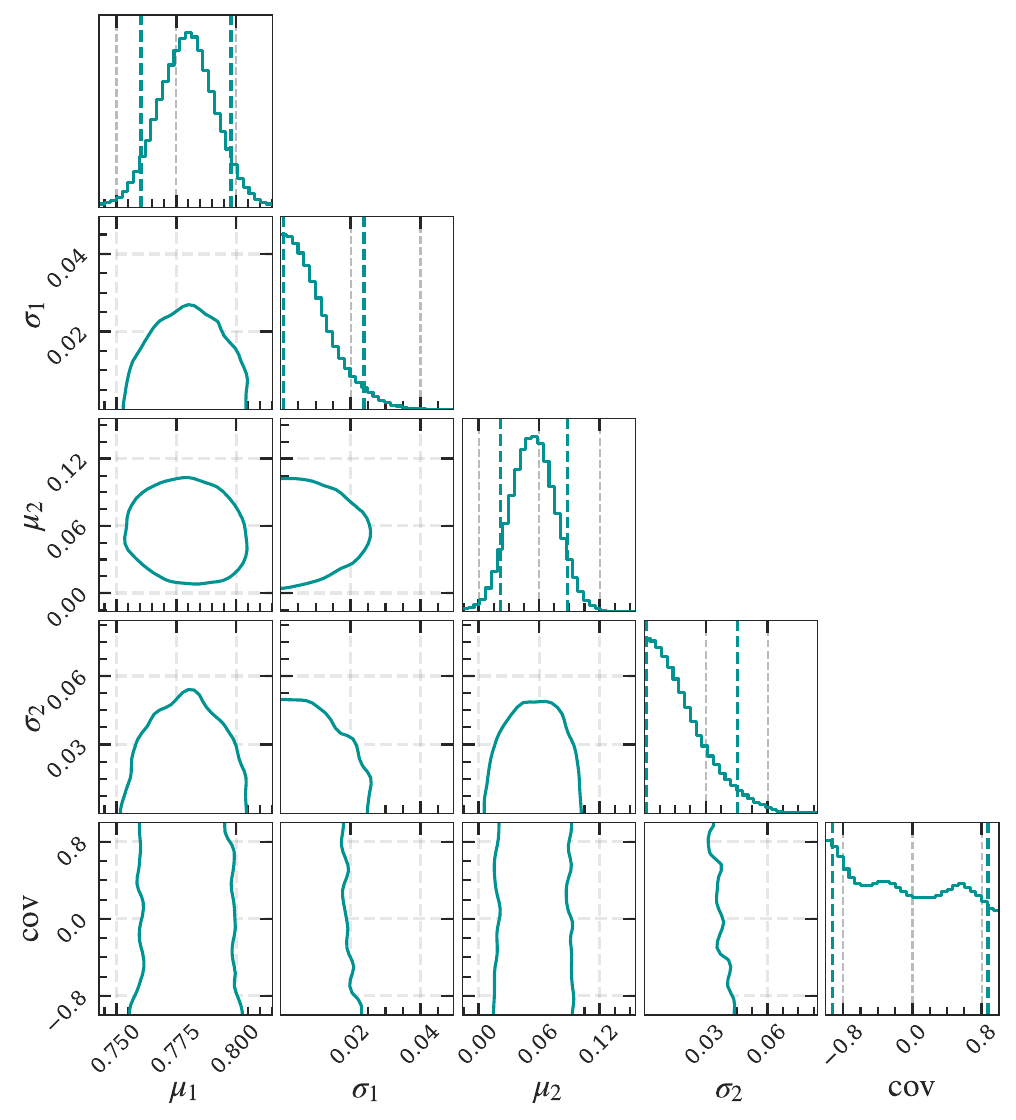}
    \caption{Posterior distribution on the hyperparameters of the fractional deviations in the complex \((220)\) \ac{qnm} frequency, obtained from the pSEOBNR analysis. The contours enclose the 90\% credible regions with \ac{gr} value being \(0\). The \(\mu_1\) hyperparameter has no support at \(0\), indicating an infinite Bayesian evidence against \ac{gr}.}
    \label{fig:mock-hierarchical}
\end{figure}

We have argued that our framework leads to less evidence against \ac{gr} because existing models impose conditions on potential \ac{gr} deviations that are too stringent to be satisfied by the data. In other words, the freedom our framework provides is necessary to accommodate the observed data within the \ac{bgr} paradigm. This should, however, not be the case if the data really satisfies the mentioned conditions.

To demonstrate this, we construct a mock event catalogue where all events share the same deviation values and compare the responses of the mixture and hierarchical Gaussian models. We create \(N\) synthetic events by drawing 1000 posterior samples from the full pSEOBNR posterior of GW191109. These, by construction, share the same underlying distribution for the deviation parameters, centred at \((\delta\omega_{220}, \delta \tau_{220})\sim (0.4, 0.4)\) as shown in Figure~\ref{fig:synthetic_events}, and doesn't include the \ac{gr} value within its 90\% contour.

Next, we compute Bayes factor against GR for each synthetic event.  Since all events have negligible support at the \ac{gr} value, the per-event log Bayes factor is strongly negative. As a consequence, the mixture model gives \(\zeta=0.01^{+0.05}_{-0.01}\), correctly identifying that the catalogue is inconsistent with \ac{gr} (Figure~\ref{fig:mock-mixture}).

The posterior on \(\boldsymbol{\mu}\) shown in Figure~\ref{fig:mock-hierarchical} is tightly constrained to \(\mu_1=0.78^{+0.01}_{-0.01}\) and \(\mu_2=0.05^{+0.02}_{-0.02}\), with no posterior support at the \ac{gr} value for \(\mu_1\). Evaluating the population-level Bayes factor gives \(\mathcal{B}^{\rm GR}_{\rm bGR}\rightarrow \infty\). This divergence is a direct consequence of the shared-mean constraint — since all events carry the same systematic offset, the posterior on \(\boldsymbol{\mu}\) tightens without bounds as \(N\) grows.

Both approaches correctly detect the displacement. However, the hierarchical framework gives an effectively infinite Bayes factor against \ac{gr}, whereas the mixture model yields a finite result --- \(\zeta\sim 0\) simply means no event in the analysed catalogue is consistent with \ac{gr}. This distinction will become increasingly consequential in the next-generation detector era, where large catalogues and high \acp{snr} will make waveform systematics a leading concern. In that regime, the mixture model naturally absorbs apparent violations driven by systematics through the Ockham penalty on \(\zeta\). The hierarchical framework, by contrast, would require identifying and excising such events through dedicated simulation studies — a procedure that rapidly becomes computationally prohibitive at the scales expected for future detectors.


\section{Conclusion}
\label{sec:conclusion}

The deluge of \ac{gw} observations over the last decade has transformed tests of \ac{gr} into a population-level endeavour. Existing approaches for combining information across events, however, make strong assumptions about how deviations from \ac{gr} should manifest in nature. In particular, they enforce an ``all-or-nothing approach'' to the number of events that deviate from \ac{gr}. Although these conditions may be appropriate in some cases, they risk producing overconfident or underconfident conclusions if nature behaves differently.

In this work, we have introduced a simple, flexible framework based on mixture models that relaxes these assumptions. By allowing only a fraction $\zeta$ of events to be inconsistent with \ac{gr}, the framework naturally accommodates a heterogeneous event population. Beyond this, it offers several practical advantages: it is robust to spurious deviations in individual events driven by waveform systematics or instrumental glitches; it can yield statistically meaningful evidence even when individual events are inconclusive on their own; and it naturally handles coexisting families of events governed by competing models.

Applying this framework to three gravitational-wave tests on O1--O4a catalogue events, we find the data consistent with all analysed events satisfying \ac{gr}. The evidence against alternative scenarios is, however, substantially weaker than that inferred under standard \ac{lvk} catalogue-combination methods --- reflecting both the additional flexibility our approach provides to accommodate potential deviations across events, and the fact that our framework tests a different hypothesis than the joint and hierarchical methods (Sec.~\ref{sec:method}). Crucially, we have shown that this flexibility does not come at a cost: when deviations from \ac{gr} are present and manifest consistently across events, our framework recovers them just as effectively as current methods, and such additional freedom would not be needed at all if deviations from \ac{gr} genuinely satisfied the restrictions imposed by current frameworks. A natural extension of this work is to account for selection effects, which we discuss in Appendix~\ref{app:selection}. In light of these results, we recommend adopting flexible frameworks of this kind, such as the one proposed here, as standard practice when testing \ac{gr} with growing gravitational-wave catalogues.

The code necessary for the analysis and for generating the plots is publicly available at ~\href{https://github.com/koustavchandra/mixturemodels}{\simpleicon{github}}.
\section*{Acknowledgments}

The authors acknowledge the use of \textsc{Claude 4.5 Sonnet}~\citep{claude-sonnet-4.5} for plotting. They also thank Giada Caneva Santoro for her comments on the manuscript.
KC thanks his co-author, JCB, for the warm hospitality in Santiago de Compostela, where the welcoming environment—and, not least, the excellent food—provided an ideal setting for kickstarting this work. He also thanks Daniel, Ana, and Samson for their company during his visit. 
JCB is supported by the Ramon y Cajal Fellowship RYC2022-036203-I,
and the research grant PID2020-118635GB-I00 from the Spain-Ministerio de Ciencia e Innovaci\'{o}n. JCB is also supported by the Grant ED431F 2025/04 of the Galician CONSELLERIA DE EDUCACION, CIENCIA, UNIVERSIDADES E FORMACION PROFESIONAL. We also acknowledge support from the European Horizon Europe staff exchange (SE) programme HORIZON-MSCA2021-SE-01 Grant No. NewFunFiCO-101086251.
IGFAE is supported by the Ayuda Maria de Maeztu CEX2023-001318-M funded by MICIU/AEI /10.13039/501100011033.
This material is based upon work supported by NSF's LIGO Laboratory, which is a major facility fully funded by the National Science Foundation. 
This paper carries LIGO document number P2600339.
\appendix

\section{Savage-Dickey density ratio in a box} 
\label{sec:savage}

In all tests of \ac{gr} considered here, \ac{gr} is nested within the \ac{bgr} model, characterised by $\boldsymbol{\delta}=0$. For these situations, the relative Bayes factor (evidence ratio) between the two models can be computed using the Savage--Dickey density ratio~\citep{Dickey:1971wlr}.
\begin{equation}\label{eq:savage-1}
    {}_k\mathcal{B}^{\rm GR}_{\rm bGR}
    = \frac{p_{\rm bGR}(\boldsymbol{\delta}=0 \mid d_k)}
           {\pi_{\rm bGR}(\boldsymbol{\delta}=0)}\,.
\end{equation}
This relates the evidence ratio to the change in probability density from prior to posterior at the \ac{gr} limit, as informed by the data. In practice, rather than evaluating densities precisely at \(\boldsymbol{\delta}=0\), the estimation is performed using a Gaussian \ac{kde} with bandwidth determined by Scott's rule. To address numerical instabilities from pointwise evaluation, we integrate over a small tolerance box of half-width \(\epsilon\) centred at zero. The numerator of Eq.~\eqref{eq:savage-1} is then given by
\begin{equation}
    \hat{p}_{\rm bGR}(\boldsymbol{\delta}=0 \mid d_k) 
    \approx \frac{1}{(2\epsilon)^2} \int^{\epsilon}_{-\epsilon} d\boldsymbol{\delta}\,\mathrm{KDE}(\boldsymbol{\delta}),
\end{equation}
while the prior density is \(\hat{\pi}_{\rm bGR}(\boldsymbol{\delta}=0) \approx (2\epsilon)^{-2}\), corresponding to a uniform distribution over the prior range. 

This procedure stabilises the estimate by averaging over a finite region and reduces sensitivity to KDE bandwidth and sampling noise, provided \(\epsilon\) remains small compared to the posterior width.

\section{How many \ac{gr}-violating events do we need to confirm true violation of \ac{gr}?}
\label{sec:violation}

An interesting and important practical question is: how many events are needed to observe \ac{gr} violations at the analysed population level? To answer this, we here derive the following analytical scaling relationship.\\

Consider a catalogue of \(N_{\rm total}\) events, where a true fraction \((1-\zeta_{\rm true})\) contains genuine \ac{gr} violations, each leading to an average single-event Bayes factor \(\langle {}_k\mathcal{B}^{\rm GR}_{\rm bGR}\rangle_{\rm violating} = 10^{-\beta}\) (disfavouring \ac{gr}, with \(\beta > 0\)). The remaining fraction \(\zeta_{\rm true}\) are \ac{gr}-consistent events leading to \(\langle {}_k\mathcal{B}^{\rm GR}_{\rm bGR}\rangle_{\rm GR} \approx 10^{\alpha}\) with \(\alpha >0\).

The population-level log-likelihood difference between the true $\zeta$ model (\(\zeta = \zeta_{\rm true}\)) and the pure-\ac{gr} hypothesis (\(\zeta = 1\)) is given by
\begin{equation}
\Delta \ln \mathfrak{L} = \ln \mathfrak{L}(\{d_k\} \mid \zeta_{\rm true}) - \ln \mathfrak{L}(\{d_k\} \mid \zeta=1)\,.
\end{equation}

To evaluate this, we decompose the catalogue into \ac{gr}-consistent and \ac{gr}-violating subpopulations. For \ac{gr}-consistent events with \({}_k\mathcal{B}^{\rm GR}_{\rm bGR} \approx 10^{\alpha}\), the contribution to the likelihood at \(\zeta = \zeta_{\rm true}\) is
\begin{equation}
\ln(\zeta_{\rm true} \cdot 10^{\alpha} + 1 - \zeta_{\rm true}) \approx \alpha \ln(10)\,,
\end{equation}
assuming \(\zeta_{\rm true} \cdot 10^{\alpha} \gg 1 - \zeta_{\rm true}\). At \(\zeta = 1\), this becomes \(\ln(10^{\alpha}) = \alpha \ln(10)\). These contributions cancel, so \ac{gr}-consistent events contribute negligibly to \(\Delta \ln \mathfrak{L}\).

For violating events with \({}_k\mathcal{B}^{\rm GR}_{\rm bGR} \approx 10^{-\beta}\), the contribution at \(\zeta = \zeta_{\rm true}\) is
\begin{equation}
\ln(\zeta_{\rm true} \cdot 10^{-\beta} + 1 - \zeta_{\rm true}) \approx \ln(1 - \zeta_{\rm true})\,,
\end{equation}
assuming \(1 - \zeta_{\rm true} \gg \zeta_{\rm true} \cdot 10^{-\beta}\). At \(\zeta = 1\), this becomes \(\ln(10^{-\beta}) = -\beta \ln(10)\). The difference per violating event is therefore
\begin{equation}
\Delta_{\rm viol} = \ln(1 - \zeta_{\rm true}) - (-\beta \ln(10)) = \beta \ln(10) + \ln(1-\zeta_{\rm true})\,.
\end{equation}

Summing over \(N_{\rm violating} = (1-\zeta_{\rm true}) N_{\rm total}\) violating events gives
\begin{equation}
\Delta \ln \mathfrak{L} \approx N_{\rm violating} \left[\beta \ln(10) + \ln(1-\zeta_{\rm true})\right]\,.
\end{equation}

For a \(3\sigma\)-equivalent detection, we require \(\Delta \ln \mathfrak{L} \gtrsim 3.3\) (corresponding to a \(\mathcal{B} \approx 27\)). This yields the detection threshold
\begin{equation}
N_{\rm violating} \gtrsim \frac{3.3}{\beta \ln(10) + \ln(1-\zeta_{\rm true})}\,.
\end{equation}

In the regime where \(\zeta_{\rm true} \to 1\) (i.e., a small violating fraction), \(\ln(1-\zeta_{\rm true}) \approx -(1-\zeta_{\rm true}) \to 0\), and this simplifies to
\begin{equation}
N_{\rm violating} \approx \frac{3.3}{\beta \ln(10)} \approx \frac{1.4}{\beta}\,.
\end{equation}

Therefore, even if individual events show only modest evidence against \ac{gr}---for instance, \(\beta \sim 0.3\) corresponding to \(\mathcal{B}^{\rm GR}_{\rm bGR} \sim 0.5\) (a factor of 2 against \ac{gr})---detecting a small subpopulation with \(1-\zeta_{\rm true} \sim 0.1\) (i.e., 10\% of events violating) requires only \(\mathcal{O}(5)\) such events. In a catalogue of \(N_{\rm total} \sim 50\), this is well within reach of current observations. Conversely, stronger per-event violations (\(\beta \sim 1\)) would be detectable with just \(\sim 1\)--\(2\) events, while weaker violations (\(\beta \sim 0.1\)) would require \(\sim 14\) events.

This scaling illustrates the power of population-level mixture modelling: by coherently combining evidence across the catalogue, systematic deviations affecting even a small fraction of events can produce conclusive population-level evidence, despite each individual event being inconclusive on its own.

\section{Comparing our method to Hierarchical inference}
\label{app:compare}

In this appendix, we make an explicit connection between mixture models and hierarchical analysis, and argue why the discrete mixture is a more natural formulation for the question at hand.

\subsection*{The hierarchical framework}

Within the hierarchical framework~\citep{Isi:2019asy}, one posits that the deviation parameters \(\boldsymbol{\delta}\) are \emph{i.i.d} random variables drawn from a population distribution governed by hyperparameters \(\boldsymbol{\Lambda}\) and evaluates the hierarchical likelihood:
\begin{equation}\label{eq:hierarchical}
        \mathfrak{L}\!\left(\{d_k\} \mid \boldsymbol{\Lambda}\right)
    = \prod_{k=1}^{N}
      \int \mathcal{L}_{\mathrm{bGR}}(d_k \mid \delta)\,
           \pi(\boldsymbol{\delta} \mid \boldsymbol{\Lambda})\, \mathrm{d}\boldsymbol{\delta}~.
\end{equation}
for a given choice of hyperprior \(\pi(\boldsymbol{\Lambda})\). Here, \(\mathcal{L}_{\mathrm{bGR}}(d_k \mid \boldsymbol{\Lambda})\) is the single-event likelihood marginalised over all \ac{gr} parameters. \Ac{gr} consistency is assessed by examining whether the posteriors \(p(\Lambda \mid \{d_k\})\) contains the \ac{gr} value \(\boldsymbol{\Lambda}_{\rm GR}\) within some predetermined credible interval. In the Gaussian hierarchical model one adopts 
\begin{equation}
    \pi(\boldsymbol{\delta}\mid \boldsymbol{\mu}, \boldsymbol{\sigma}) \sim \mathcal{N}(\boldsymbol{\mu}, \boldsymbol{\sigma})
\end{equation}
so that $\boldsymbol{\Lambda} = (\boldsymbol{\mu}, \boldsymbol{\sigma})$ and \ac{gr} corresponds to \(\boldsymbol{\mu}=0, \boldsymbol{\sigma}=0\).





\subsection*{Hierarchical inference as a restricted mixture model}
\label{app:pop-hier}

The traditional hierarchical framework can be understood as a heavily constrained version of our flexible mixture model. 

First, the hierarchical model strictly imposes the condition $\zeta = 0$, forcing all events in the catalog to be described entirely by the broader, continuous \ac{bgr} population model. Second, our unconstrained mixture model utilizes independent single-event Bayes factors obtained by letting the deviation parameters vary completely freely according to an individual event prior $\pi(\boldsymbol{\delta})$. Instead, the hierarchical model imposes that single-event deviations must be drawn from a restricted prior for the parameters $\delta$ given by the common population distribution $\pi(\boldsymbol{\delta} \mid \boldsymbol{\Lambda}) \sim \mathcal{N}(\boldsymbol{\mu}, \boldsymbol{\sigma})$ governed by global hyper-parameters $\boldsymbol{\Lambda} = (\boldsymbol{\mu}, \boldsymbol{\sigma})$. 

Under these two restrictions ($\zeta = 0$ and a constrained population prior), the catalog likelihood in Eq.~\eqref{eq:analysed-likelihood} simplifies to the traditional hierarchical form:

\begin{equation}
    \mathfrak{L}(\{d_k\} \mid \boldsymbol{\Lambda}) = \prod_{k=1}^{N} \mathcal{Z}_{\mathrm{bGR}}(d_k|\boldsymbol{\Lambda}). 
\end{equation}

Here, $\mathcal{Z}_{\mathrm{bGR}}(d_k|\boldsymbol{\Lambda})$ is the single-event marginal likelihood obtained under a particular choice of $\boldsymbol{\Lambda}$, given by

\begin{equation}
\mathcal{Z}_{\mathrm{bGR}}(d_k|\boldsymbol{\Lambda}) = \int \mathcal{L}(d_k|\boldsymbol{\delta})\pi(\boldsymbol{\delta} \mid \boldsymbol{\Lambda}) d \boldsymbol{\delta}.
\end{equation}

in contrast with the mixture model single-event evidences given by

\begin{equation}
\mathcal{Z}_{\mathrm{bGR}}(d_k) = \int \mathcal{L}(d_k|\boldsymbol{\delta})\pi(\boldsymbol{\delta}) d \boldsymbol{\delta}.
\end{equation}

Thus, the hierarchical likelihood can be understood as the limit $\zeta = 0$ of our mixture model likelihood, where individual events are analyzed under a common prior determined by the population hyper-parameters $\Lambda$. A similar derivation can be found in \citep{Lorenzo-Medina:2025ylc}.\\ 

\section{Estimating the Bayes Factor for Hierarchical Analysis}
\label{app:hierarchical_bayes_factors}

To assess population-level consistency with \ac{gr}, we evaluate the Bayes factor between the \ac{gr} hypothesis, corresponding to hyperparameters fixed at \(\boldsymbol{\Lambda}_{\rm GR}\), and the more general hierarchical model in which \(\boldsymbol{\Lambda}\) is allowed to vary. Since \ac{gr} is nested within the hierarchical model, this Bayes factor can be computed using the Savage--Dickey density ratio:
\begin{equation}
    \mathfrak{B}^{\mathrm{GR}}_{\mathrm{bGR}}
    = \frac{p(\boldsymbol{\Lambda} = \boldsymbol{\Lambda}_{\rm GR} \mid \{d_k\})}
       {\pi(\boldsymbol{\Lambda} = \boldsymbol{\Lambda}_{\rm GR})}\,.
\end{equation}

Like before we evaluate it over a small tolerance region centred around \(\boldsymbol{\Lambda}_{\rm GR}\):
\begin{equation}
\mathfrak{B}^{\mathrm{GR}}_{\mathrm{bGR}}
\approx
\frac{\int_{\Box \epsilon} p(\boldsymbol{\Lambda} \mid \{d_k\})\, d\boldsymbol{\Lambda}}
     {\int_{\Box \epsilon} \pi(\boldsymbol{\Lambda})\, d\boldsymbol{\Lambda}}\,.
\end{equation}
For parameters with a natural zero point (e.g., population means), we adopt symmetric bounds \([-\epsilon, \epsilon]\), while for parameters constrained to be positive (e.g., population standard deviations), we use one-sided intervals \([0, \epsilon]\). 

For uniform hyperpriors, the prior mass within \(\Box \epsilon\) is analytic. The posterior mass is estimated from samples using a Gaussian Mixture model and verified using normalising flow. This is because kernel density estimates become unreliable in higher dimensions.

\section{Extending mixture models to astrophysical population level including selection effects}
\label{app:selection}

The mixture model framework developed in the main text assumes that the detection probability is independent of which hypothesis (\ac{gr} versus beyond-\ac{gr}) generated the signal. In practice, waveform deviations can affect detectability, introducing a selection bias that must be accounted for when inferring the population-level \ac{gr} fraction \(\zeta\).

\subsection*{Incorporating Selection Effects}

If the detection probability differs between the two hypotheses, the population-level likelihood gains a correction. Suppose \(\alpha_{\mathcal{H}_0}\) and \(\alpha_{\mathcal{H}_1}\) denote the detection probabilities under \(\mathcal{H}_0\) (\ac{gr}) and \(\mathcal{H}_1\) (beyond-\ac{gr}), respectively. Then for a catalogue drawn from a mixture with \ac{gr} fraction \(\zeta\), the effective detection probability is
\begin{equation}
    \alpha(\zeta) = \zeta \, \alpha_{\mathcal{H}_0} + (1-\zeta) \, \alpha_{\mathcal{H}_1}\,.
\end{equation}

Each detected event is drawn from the detectable population, not the underlying astrophysical population. The selection-corrected likelihood becomes
\begin{equation}
    \ln \mathfrak{L}^{\rm select}(\{d_k\} \mid \zeta) = \ln \mathfrak{L}(\{d_k\} \mid \zeta) - N \ln \alpha(\zeta)\,,
\end{equation}
where \(N\) is the number of detected events and \(\ln \mathfrak{L}(\{d_k\} \mid \zeta)\) is the likelihood given in the main text. The second term penalises models that predict lower overall detection rates, ensuring that the inferred \(\zeta\) accounts for differential selection.

\subsection*{Estimating Detection Probabilities}

Computing \(\alpha_{\mathcal{H}_0}\) and \(\alpha_{\mathcal{H}_1}\) requires adding signals into detector noise and determining the fraction recovered above threshold. For \(\mathcal{H}_0\), this is standard: one adds \ac{gr} waveforms and measures the detection efficiency as a function of source parameters (masses, spins, distance, sky location). 

For \(\mathcal{H}_1\), the situation is more subtle. Parameterised tests do not commit to a specific beyond-\ac{gr} theory, so there is no unique alternative waveform family to simulate and analyse. However, one can adopt the following pragmatic approach:

\paragraph{Approximation 1: Negligible selection effect.} For most events, waveform deviations parameterised by \(\boldsymbol{\delta}\)typically produce small fractional changes in amplitude and phase. If these changes do not significantly affect the matched-filter \ac{snr}, the detection probability remains approximately unchanged:
\begin{equation}
    \alpha_{\mathcal{H}_1} \approx \alpha_{\mathcal{H}_0}\,.
\end{equation}
In this regime, \(\alpha(\zeta) \approx \alpha_{\mathcal{H}_0}\) is independent of \(\zeta\), and the selection correction vanishes. This approximation is justified when deviations are small, as is typical in current analyses. However, one should keep in mind that detectability depends not only on the matched-filter \ac{snr} but also on the output of the various signal-noise discriminators, which are used to augment the matched-filter \ac{snr} so as to separate the signal and noise population. Given that the templates used in searches are \ac{gr}-faithful, this can have a non-negligible impact, particularly for extremely loud events. 

\paragraph{Approximation 2: Empirical selection estimate.}
If deviations are expected to be non-negligible, one can estimate \(\alpha_{\mathcal{H}_1}\) by adding morphed waveforms \(h_{\rm bGR} = h_{\rm GR}(1 + \Delta h(\boldsymbol{\delta}))\) with \(\boldsymbol{\delta}\) drawn from the prior \(\pi_{\rm bGR}(\boldsymbol{\delta})\). The detection efficiency is then averaged over the prior:
\begin{equation}
    \alpha_{\mathcal{H}_1} = \int p_{\rm det}(\boldsymbol{\theta}, \boldsymbol{\delta}) \, \pi_{\rm pop}(\boldsymbol{\theta}) \, \pi_{\rm bGR}(\boldsymbol{\delta}) \, d\boldsymbol{\theta} \, d\boldsymbol{\delta}\,,
\end{equation}
where \(p_{\rm det}(\boldsymbol{\theta}, \boldsymbol{\delta})\) is the detection probability for a signal with \ac{gr} parameters \(\boldsymbol{\theta}\) and deviations \(\boldsymbol{\delta}\), and \(\pi_{\rm pop}(\boldsymbol{\theta})\) is the astrophysical population distribution. This requires adding a representative set of morphed waveforms to the detector noise and recovering them via searches. Alternatively, one can calculate \(\bar{\cal F}\) for \ac{bgr} waveforms with a \ac{gr} faithful template bank and use it to estimate the matched-filter \ac{snr} and \(\chi_r^2\) and consequently predict the detection probability of that \ac{bgr} signal~\citep{Allen:2004gu, Harry:2016ijz}.

\subsection*{Impact on the Analysis}

In the main text, we adopt Approximation 1 and set \(\alpha_{\mathcal{H}_0} = \alpha_{\mathcal{H}_1}\), so that selection effects drop out. This is well-justified for the analysed events, which are loud (\(\rho \gtrsim 10\)) and where deviation parameters are constrained to relatively small values by the data. For future catalogues with marginal detections or if strong deviations are suspected, Approximations 2 needs to be implemented.

We note that the selection-corrected posterior \(\mathfrak{p}^{\rm select}(\zeta \mid \{d_k\})\) is straightforward to compute given \(\alpha_{\mathcal{H}_0}\) and \(\alpha_{\mathcal{H}_1}\): one simply evaluates \(\ln \mathfrak{L}^{\rm select}(\zeta)\) in place of \(\ln \mathfrak{L}(\zeta)\) when computing the posterior.

\bibliography{aapmsamp}
\end{document}